\input harvmac
\input amssym

\def\unit{\relax{\rm 1\kern-.26em I}}
\def\nada{\relax{\rm 0\kern-.30em l}}
\def\tilde{\widetilde}

\def\alphadot{{\dot \alpha}}



\def\det{{\rm det}}

\noblackbox
\def\IL{\relax{\rm I\kern-.18em L}}
\def\IH{\relax{\rm I\kern-.18em H}}
\def\IR{\relax{\rm I\kern-.18em R}}
\def\IC{\relax\hbox{$\inbar\kern-.3em{\rm C}$}}
\def\IZ{\relax\ifmmode\mathchoice
{\hbox{\cmss Z\kern-.4em Z}}{\hbox{\cmss Z\kern-.4em Z}}
{\lower.9pt\hbox{\cmsss Z\kern-.4em Z}} {\lower1.2pt\hbox{\cmsss
Z\kern-.4em Z}}\else{\cmss Z\kern-.4em Z}\fi}

\def\CN {{\cal N}}
\def\CR {{\cal R}}

\def\CJ {{\cal J}}

\def\CL {{\cal L}}

\def\CO {{\cal O}}


\def\CN {{\cal N}}

\def\CO {{\cal O}}

\def\det{{\rm det}}
\def\Tr{{\rm Tr}}

\font\manual=manfnt \def\dbend{\lower3.5pt\hbox{\manual\char127}}

\def\IZ{\relax\ifmmode\mathchoice
{\hbox{\cmss Z\kern-.4em Z}}{\hbox{\cmss Z\kern-.4em Z}}
{\lower.9pt\hbox{\cmsss Z\kern-.4em Z}} {\lower1.2pt\hbox{\cmsss
Z\kern-.4em Z}}\else{\cmss Z\kern-.4em Z}\fi}
\def\half {{1\over 2}}

\def\bar{\overline}

\def\rt2{\sqrt{2}}
\def\irt2{{1\over\sqrt{2}}}

\def\slashchar#1{\setbox0=\hbox{$#1$}           
   \dimen0=\wd0                                 
   \setbox1=\hbox{/} \dimen1=\wd1               
   \ifdim\dimen0>\dimen1                        
      \rlap{\hbox to \dimen0{\hfil/\hfil}}      
      #1                                        
   \else                                        
      \rlap{\hbox to \dimen1{\hfil$#1$\hfil}}   
      /                                         
   \fi}

\def\foursqr#1#2{{\vcenter{\vbox{
    \hrule height.#2pt
    \hbox{\vrule width.#2pt height#1pt \kern#1pt
    \vrule width.#2pt}
    \hrule height.#2pt
    \hrule height.#2pt
    \hbox{\vrule width.#2pt height#1pt \kern#1pt
    \vrule width.#2pt}
    \hrule height.#2pt
        \hrule height.#2pt
    \hbox{\vrule width.#2pt height#1pt \kern#1pt
    \vrule width.#2pt}
    \hrule height.#2pt
        \hrule height.#2pt
    \hbox{\vrule width.#2pt height#1pt \kern#1pt
    \vrule width.#2pt}
    \hrule height.#2pt}}}}
\def\psqr#1#2{{\vcenter{\vbox{\hrule height.#2pt
    \hbox{\vrule width.#2pt height#1pt \kern#1pt
    \vrule width.#2pt}
    \hrule height.#2pt \hrule height.#2pt
    \hbox{\vrule width.#2pt height#1pt \kern#1pt
    \vrule width.#2pt}
    \hrule height.#2pt}}}}
\def\sqr#1#2{{\vcenter{\vbox{\hrule height.#2pt
    \hbox{\vrule width.#2pt height#1pt \kern#1pt
    \vrule width.#2pt}
    \hrule height.#2pt}}}}

\def\figin{\epsfcheck\figin}\def\figins{\epsfcheck\figins}
\def\epsfcheck{\ifx\epsfbox\UnDeFiNeD
\message{(NO epsf.tex, FIGURES WILL BE IGNORED)}
\gdef\figin##1{\vskip2in}\gdef\figins##1{\hskip.5in}
\else\message{(FIGURES WILL BE INCLUDED)}%
\gdef\figin##1{##1}\gdef\figins##1{##1}\fi}
\def\DefWarn#1{}
\def\figinsert{\goodbreak\midinsert}
\def\ifig#1#2#3{\DefWarn#1\xdef#1{fig.~\the\figno}
\writedef{#1\leftbracket fig.\noexpand~\the\figno}%
\figinsert\figin{\centerline{#3}}\medskip\centerline{\vbox{\baselineskip12pt
\advance\hsize by -1truein\noindent\footnotefont{\bf
Fig.~\the\figno:\ } \it#2}}
\bigskip\endinsert\global\advance\figno by1}

\lref\FerraraPZ{
  S.~Ferrara, B.~Zumino,
  ``Transformation Properties Of The Supercurrent,''
  Nucl.\ Phys.\  B {\bf 87}, 207 (1975).
}

\lref\KomargodskiRB{
  Z.~Komargodski, N.~Seiberg,
  ``Comments on Supercurrent Multiplets, Supersymmetric Field Theories and
  Supergravity,''
  JHEP {\bf 1007}, 017 (2010)
  [arXiv:1002.2228 [hep-th]].
}

\lref\SeibergPQ{
  N.~Seiberg,
  ``Electric - magnetic duality in supersymmetric nonAbelian gauge theories,''
  Nucl.\ Phys.\  B {\bf 435}, 129 (1995)
  [arXiv:hep-th/9411149].
}

\lref\KomargodskiPC{
  Z.~Komargodski, N.~Seiberg,
  ``Comments on the Fayet-Iliopoulos Term in Field Theory and Supergravity,''
  JHEP {\bf 0906}, 007 (2009)
  [arXiv:0904.1159 [hep-th]].
}

\lref\ArkaniHamedWC{
  N.~Arkani-Hamed, R.~Rattazzi,
  ``Exact results for nonholomorphic masses in softly broken supersymmetric gauge theories,''
Phys.\ Lett.\  {\bf B454}, 290-296 (1999).
[hep-th/9804068].
}

\lref\KutasovVE{
  D.~Kutasov,
  ``A Comment on duality in N=1 supersymmetric nonAbelian gauge theories,''
Phys.\ Lett.\  {\bf B351}, 230-234 (1995).
[hep-th/9503086].
}

\lref\KutasovNP{
  D.~Kutasov, A.~Schwimmer,
  ``On duality in supersymmetric Yang-Mills theory,''
Phys.\ Lett.\  {\bf B354}, 315-321 (1995).
[hep-th/9505004].
}

\lref\KutasovSS{
  D.~Kutasov, A.~Schwimmer, N.~Seiberg,
  ``Chiral rings, singularity theory and electric - magnetic duality,''
Nucl.\ Phys.\  {\bf B459}, 455-496 (1996).
[hep-th/9510222].
}

\lref\AbelTX{
  S.~Abel, V.~V.~Khoze,
  ``Direct Mediation, Duality and Unification,''
JHEP {\bf 0811}, 024 (2008). [arXiv:0809.5262 [hep-ph]].
}

\lref\LutyQC{
  M.~A.~Luty, R.~Rattazzi,
  ``Soft supersymmetry breaking in deformed moduli spaces, conformal theories, and N=2 Yang-Mills theory,''
JHEP {\bf 9911}, 001 (1999).
[hep-th/9908085].
}

\lref\KomargodskiRZ{
  Z.~Komargodski, N.~Seiberg,
  ``From Linear SUSY to Constrained Superfields,''
JHEP {\bf 0909}, 066 (2009).
[arXiv:0907.2441 [hep-th]].
}

\lref\IntriligatorAU{
  K.~A.~Intriligator, N.~Seiberg,
  ``Lectures on supersymmetric gauge theories and electric - magnetic duality,''
Nucl.\ Phys.\ Proc.\ Suppl.\  {\bf 45BC}, 1-28 (1996).
[hep-th/9509066].
}

\lref\WittenVV{
  E.~Witten,
``Current Algebra Theorems for the U(1) Goldstone Boson,''
  Nucl.\ Phys.\  B {\bf 156}, 269 (1979).
}

\lref\SeibergBZ{
  N.~Seiberg,
  ``Exact results on the space of vacua of four-dimensional SUSY gauge theories,''
Phys.\ Rev.\  {\bf D49}, 6857-6863 (1994). [hep-th/9402044].
}

\lref\IntriligatorAU{
  K.~A.~Intriligator, N.~Seiberg,
  ``Lectures on supersymmetric gauge theories and electric - magnetic duality,''
Nucl.\ Phys.\ Proc.\ Suppl.\  {\bf 45BC}, 1-28 (1996).
[hep-th/9509066].
}

\lref\IntriligatorJJ{
  K.~A.~Intriligator, B.~Wecht,
  ``The Exact superconformal R symmetry maximizes a,''
Nucl.\ Phys.\  {\bf B667}, 183-200 (2003). [hep-th/0304128].
}

\lref\SeibergPQ{
  N.~Seiberg,
  ``Electric - magnetic duality in supersymmetric nonAbelian gauge theories,''
Nucl.\ Phys.\  {\bf B435}, 129-146 (1995). [hep-th/9411149].
}

\lref\ManoharIY{
  A.~V.~Manohar,
  ``Wess-Zumino terms in supersymmetric gauge theories,''
Phys.\ Rev.\ Lett.\  {\bf 81}, 1558-1561 (1998). [hep-th/9805144].
}

\lref\SeibergBZ{
  N.~Seiberg,
  ``Exact results on the space of vacua of four-dimensional SUSY gauge theories,''
Phys.\ Rev.\  {\bf D49}, 6857-6863 (1994). [hep-th/9402044].
}

\lref\ChengXG{
  H.~-C.~Cheng, Y.~Shadmi,
  ``Duality in the presence of supersymmetry breaking,''
Nucl.\ Phys.\  {\bf B531}, 125-150 (1998). [hep-th/9801146].
}

\lref\DineZA{
  M.~Dine, W.~Fischler, M.~Srednicki,
  ``Supersymmetric Technicolor,''
Nucl.\ Phys.\  {\bf B189}, 575-593 (1981).
}

\lref\DobrescuGZ{
  B.~A.~Dobrescu,
  ``Fermion masses without Higgs: A Supersymmetric technicolor model,''
Nucl.\ Phys.\  {\bf B449}, 462-482 (1995). [hep-ph/9504399].
}

\lref\LutyFJ{
  M.~A.~Luty, J.~Terning, A.~K.~Grant,
  ``Electroweak symmetry breaking by strong supersymmetric dynamics at the TeV scale,''
Phys.\ Rev.\  {\bf D63}, 075001 (2001). [hep-ph/0006224].
}

\lref\FukushimaPM{
  H.~Fukushima, R.~Kitano, M.~Yamaguchi,
  ``SuperTopcolor,''
JHEP {\bf 1101}, 111 (2011). [arXiv:1012.5394 [hep-ph]].
}

\lref\KutasovIY{
  D.~Kutasov, A.~Parnachev, D.~A.~Sahakyan,
 ``Central charges and U(1)(R) symmetries in N=1 superYang-Mills,''
JHEP {\bf 0311}, 013 (2003). [hep-th/0308071].
}

\lref\IntriligatorMI{
  K.~A.~Intriligator, B.~Wecht,
  ``RG fixed points and flows in SQCD with adjoints,''
Nucl.\ Phys.\  {\bf B677}, 223-272 (2004). [hep-th/0309201].
}

\lref\DumitrescuCA{
  T.~T.~Dumitrescu, Z.~Komargodski, M.~Sudano,
  ``Global Symmetries and D-Terms in Supersymmetric Field Theories,''
JHEP {\bf 1011}, 052 (2010). [arXiv:1007.5352 [hep-th]].
}

\lref\ThomasNati{
  T.~T.~Dumitrescu ,  N.~Seiberg,
  To Appear
}

\lref\AharonyZH{
  O.~Aharony, J.~Sonnenschein, M.~E.~Peskin, S.~Yankielowicz,
  ``Exotic nonsupersymmetric gauge dynamics from supersymmetric QCD,''
Phys.\ Rev.\  {\bf D52}, 6157-6174 (1995). [hep-th/9507013].
}

\lref\WittenTW{
  E.~Witten,
  ``Global Aspects of Current Algebra,''
Nucl.\ Phys.\  {\bf B223}, 422-432 (1983).
}

\lref\SundrumGV{
  R.~Sundrum,
  ``SUSY Splits, But Then Returns,''
JHEP {\bf 1101}, 062 (2011). [arXiv:0909.5430 [hep-th]].
}

\lref\AntoniadisGN{
  I.~Antoniadis, M.~Buican,
  ``On R-symmetric Fixed Points and Superconformality,''
[arXiv:1102.2294 [hep-th]].
}

\lref\PolchinskiDY{
  J.~Polchinski,
  ``Scale And Conformal Invariance In Quantum Field Theory,''
Nucl.\ Phys.\  {\bf B303}, 226 (1988).
}

\lref\GatesNR{
  S.~J.~Gates, M.~T.~Grisaru, M.~Rocek, W.~Siegel,
  ``Superspace Or One Thousand and One Lessons in Supersymmetry,''
Front.\ Phys.\  {\bf 58}, 1-548 (1983).
[hep-th/0108200].
}

\lref\MagroAJ{
  M.~Magro, I.~Sachs, S.~Wolf,
  ``Superfield Noether procedure,''
Annals Phys.\  {\bf 298}, 123-166 (2002).
[hep-th/0110131].
}

\lref\KuzenkoAM{
  S.~M.~Kuzenko,
  ``Variant supercurrent multiplets,''
JHEP {\bf 1004}, 022 (2010).
[arXiv:1002.4932 [hep-th]].
}

\lref\KuzenkoNI{
  S.~M.~Kuzenko,
  ``Variant supercurrents and Noether procedure,''
Eur.\ Phys.\ J.\  {\bf C71}, 1513 (2011).
[arXiv:1008.1877 [hep-th]].
}

\lref\AntoniadisNJ{
  I.~Antoniadis, M.~Buican,
  ``Goldstinos, Supercurrents and Metastable SUSY Breaking in N=2 Supersymmetric Gauge Theories,''
JHEP {\bf 1104}, 101 (2011).
[arXiv:1005.3012 [hep-th]].
}

\lref\ButterSC{
  D.~Butter, S.~M.~Kuzenko,
  ``N=2 supergravity and supercurrents,''
JHEP {\bf 1012}, 080 (2010).
[arXiv:1011.0339 [hep-th]].
}

\lref\YonekuraMC{
  K.~Yonekura,
  ``Notes on Operator Equations of Supercurrent Multiplets and Anomaly Puzzle in Supersymmetric Field Theories,''
JHEP {\bf 1009}, 049 (2010).
[arXiv:1004.1296 [hep-th]].
}

\lref\ArkaniHamedMJ{
  N.~Arkani-Hamed, H.~Murayama,
  ``Holomorphy, rescaling anomalies and exact beta functions in supersymmetric gauge theories,''
JHEP {\bf 0006}, 030 (2000).
[hep-th/9707133].
}

\lref\HuangTN{
  X.~Huang, L.~Parker,
  ``Clarifying Some Remaining Questions in the Anomaly Puzzle,''
Eur.\ Phys.\ J.\  {\bf C71}, 1570 (2011).
[arXiv:1001.2364 [hep-th]].
}

\lref\ShifmanZI{
  M.~A.~Shifman, A.~I.~Vainshtein,
  ``Solution of the Anomaly Puzzle in SUSY Gauge Theories and the Wilson Operator Expansion,''
Nucl.\ Phys.\  {\bf B277}, 456 (1986).
}

\lref\GreenDA{
  D.~Green, Z.~Komargodski, N.~Seiberg, Y.~Tachikawa, B.~Wecht,
  ``Exactly Marginal Deformations and Global Symmetries,''
JHEP {\bf 1006}, 106 (2010). [arXiv:1005.3546 [hep-th]].
}

\lref\DorigoniRA{
  D.~Dorigoni, V.~S.~Rychkov,
  ``Scale Invariance + Unitarity $=>$ Conformal Invariance?,''
[arXiv:0910.1087 [hep-th]].
}

\lref\KarchQA{
  A.~Karch, T.~Kobayashi, J.~Kubo, G.~Zoupanos,
  ``Infrared behavior of softly broken SQCD and its dual,''
Phys.\ Lett.\  {\bf B441}, 235-242 (1998). [hep-th/9808178].
}

\lref\KobayashiWK{
  T.~Kobayashi, K.~Yoshioka,
  ``New RG invariants of soft supersymmetry breaking parameters,''
Phys.\ Lett.\  {\bf B486}, 223-227 (2000). [hep-ph/0004175].
}

\lref\Future{
  In Progress
}

\lref\NelsonMQ{
  A.~E.~Nelson, M.~J.~Strassler,
  ``Exact results for supersymmetric renormalization and the supersymmetric flavor problem,''
JHEP {\bf 0207}, 021 (2002). [hep-ph/0104051].
}

\lref\DienesTD{
  K.~R.~Dienes, B.~Thomas,
  ``On the Inconsistency of Fayet-Iliopoulos Terms in Supergravity Theories,''
Phys.\ Rev.\  {\bf D81}, 065023 (2010). [arXiv:0911.0677
[hep-th]].
}

\lref\DumitrescuCA{
  T.~T.~Dumitrescu, Z.~Komargodski, M.~Sudano,
  ``Global Symmetries and D-Terms in Supersymmetric Field Theories,''
JHEP {\bf 1011}, 052 (2010). [arXiv:1007.5352 [hep-th]].
}

\lref\KolZT{
  B.~Kol,
  ``On conformal deformations,''
JHEP {\bf 0209}, 046 (2002). [hep-th/0205141].
}

\lref\EvansIA{
  N.~J.~Evans, S.~D.~H.~Hsu, M.~Schwetz,
  ``Exact results in softly broken supersymmetric models,''
Phys.\ Lett.\  {\bf B355}, 475-480 (1995).
[hep-th/9503186].
}

\lref\EvansRV{
  N.~J.~Evans, S.~D.~H.~Hsu, M.~Schwetz, S.~B.~Selipsky,
  ``Exact results and soft breaking masses in supersymmetric gauge theory,''
Nucl.\ Phys.\  {\bf B456}, 205-218 (1995).
[hep-th/9508002].
}

\lref\ShifmanKA{
  M.~Shifman, A.~Yung,
  ``Non-Abelian Duality and Confinement: from N=2 to N=1 Supersymmetric QCD,''
[arXiv:1103.3471 [hep-th]].
}

\lref\ShifmanXC{
  M.~Shifman, W.~Vinci, A.~Yung,
  ``Effective World-Sheet Theory for Non-Abelian Semilocal Strings in N = 2 Supersymmetric QCD,''
[arXiv:1104.2077 [hep-th]].
}

\lref\KomargodskiMC{
  Z.~Komargodski,
  ``Vector Mesons and an Interpretation of Seiberg Duality,''
JHEP {\bf 1102}, 019 (2011). [arXiv:1010.4105 [hep-th]].
}

\lref\EtoPJ{
  M.~Eto, T.~Fujimori, M.~Nitta, K.~Ohashi, N.~Sakai,
  ``Dynamics of Non-Abelian Vortices,''
[arXiv:1105.1547 [hep-th]].
}

\rightline{CERN-PH-TH/2011-112}
\Title{
} {\vbox{\centerline{ Mapping Anomalous Currents in Supersymmetric
Dualities } }}
\medskip

\centerline{\it Steven Abel,$^{\clubsuit\diamondsuit}$ Matthew
Buican,$^\diamondsuit$ and Zohar
Komargodski$^{\spadesuit\heartsuit}$}
\bigskip
\centerline{$^\clubsuit$ Institute for Particle Physics Phenomenology, Durham University, UK} \centerline{$^\diamondsuit$ Department of Physics, CERN Theory Division, CH-1211 Geneva 23, Switzerland}
 \centerline{$^\spadesuit$
Institute for Advanced Study, Princeton, NJ 08540, USA}
\centerline{$^\heartsuit$ Weizmann Institute of Science, Rehovot
76100, Israel}

\smallskip

\vglue .3cm

\bigskip
\noindent

In many strongly-coupled systems, the infrared dynamics is
described by different degrees of freedom from the ultraviolet. It
is then natural to ask how operators written in terms of the
microscopic variables are mapped to operators composed of the
macroscopic ones. Certain types of operators, like conserved
currents, are simple to map, and in supersymmetric theories one
can also follow the chiral ring. In this note, we consider
supersymmetric theories and extend the mapping to anomalous
currents (and gaugino bilinears). Our technique is completely
independent of subtleties associated with the renormalization
group, thereby shedding new light on previous approaches to the
problem. We demonstrate the UV/IR mapping in several examples with
different types of dynamics, emphasizing the uniformity and
simplicity of the approach. Natural applications of these ideas
include the effects of soft breaking on the dynamics of various
theories and new models of electroweak symmetry breaking.

\Date{May 2011}

\newsec{Introduction}

At low energies, many strongly-coupled field theories can be described in terms of emergent degrees of freedom---often markedly different from those used to define the theory at short distances. The most well-known example where this phenomenon occurs is QCD, which, in the chiral limit, flows from a theory described purely in terms of fermions and gauge fields to a free theory of massless pions.

Given this picture, an important question that arises is how to express long-distance correlation functions, written in terms of the fundamental quarks and color gauge fields, as correlation functions written in terms of mesons. (Of course, in order for this question to be well-defined, one must only consider gauge-invariant correlation functions in the UV.)

For instance, one may consider correlation functions of conserved
currents, which in QCD are associated with the symmetries
$SU(N_f)_L\times SU(N_f)_R\times U(1)_B$ and attempt to rewrite
the corresponding conserved currents in terms of the pions. This
procedure is fairly straightforward for the non-Abelian currents,
but some interesting complications arise for the baryonic current
(for a general treatment see~\WittenTW).

Another interesting set of  quark bilinears in QCD are the
$\psi\tilde \psi$ operators. In this case, one can use an
$SU(N_f)_L\times SU(N_f)_R$ spurion analysis and find that they
map to $U=e^{i\pi^aT^a}$. Unfortunately, the coefficient in this
mapping is incalculable. One can also consider the quark bilinear
corresponding to the anomalous axial current, $U(1)_A$. However,
we are not aware of any systematic procedure of mapping this
operator to the IR.\foot{The situation might be better in the
large $N$ limit of QCD; there one can imagine including the light
$\eta'$ particle~\WittenVV.}

In  this paper we will discuss related questions in the context of supersymmetric (SUSY) theories. In SUSY theories it is relatively straightforward to follow the flows of two broad classes of operators---elements of the chiral ring and the (non-chiral) conserved current multiplets. The mapping of the conserved currents follows the same rules as in non-SUSY theories. We use either 't-Hooft anomaly matching or Goldstone's theorem to realize various conserved currents in the IR.\foot{Complications, as for baryon number in QCD, can arise too, although they do not arise in the simplest examples. See~\ManoharIY\ for interesting discussions of closely related matters.} The mapping of the chiral ring is, of course, possible due to the strong constraints imposed by holomorphy.

Generalizing the above ideas to non-conserved currents (and
objects that vanish in the chiral ring) is more difficult.
However, understanding their flow is crucial for many
applications, such as the mapping of soft non-holomorphic mass
terms, which, at weak coupling, can be thought of as the lowest
components of (non)-conserved current multiplets. Studying these
questions is the chief goal of this paper.

The main utility of supersymmetry in this context is as follows. Consider a current broken explicitly by an anomaly. It satisfies the Adler-Bardeen equation (it is actually not important for us to work in a scheme where the anomaly is one-loop exact)
\eqn\abj{
\del^{\mu} j_{\mu}\sim F\tilde F~.
}
However, supersymmetry relates $F\tilde F$ with $F^2$ since they together form the complex $\theta^2$ component of $W_{\alpha}^2$. The final crucial ingredient is that $F^2$ is related to the stress tensor via the usual trace anomaly
\eqn\traceanomaly
{
T_\mu^\mu\sim F^2.
}
Even if the theory goes through strong coupling, the {\it conserved} energy-momentum tensor is known at the end points of the flow (as long as there is a description in terms of weakly-coupled degrees of freedom there). From this discussion, we see that we can follow $F\tilde F$ and learn something about the flow of anomalous currents.

The problem can be simplified even further if there is a conserved
$R$-symmetry. Indeed, the corresponding $R$-current is related to
the energy-momentum tensor by SUSY. Being a conserved current, the
$R$-current is easily followed along the flow. Therefore, in a
heuristic sense, supersymmetry extends the simplicity of the flow
of the conserved $R$-current to the flow of the anomalous (non-$R$)
current.

The question of following non-holomorphic operators, at least in
the guise of soft-SUSY breaking masses, to the IR is not new.
Indeed, there is a significant literature on the subject---see e.g. \refs{\EvansIA\AharonyZH\EvansRV\ChengXG\ArkaniHamedWC\KarchQA\KobayashiWK-\NelsonMQ} and references therein---as well as a framework for understanding many aspects of this problem \LutyQC: our paper is most closely related to this latter work. While our paper has many new concrete
results, its main purpose is to advocate several new points of
view on the subject. These perspectives allow us to solve the
problem while completely avoiding discussions of subtleties
associated with RG invariant versus RG non-invariant (i.e.
scheme-independent versus scheme-dependent) quantities. Most
importantly, our tools provide us with general results that are
valid uniformly in all theories considered, including ones that
have not been addressed before.

The plan of this paper is as follows. In section~2 we describe in
detail the procedure outlined above. In sections~3, 4, and 5 we
discuss three examples in which one can map some non-chiral
operators to a weakly-coupled dual description. These examples
demonstrate slightly different aspects and nuances of the general
procedure. In section~6 we consider theories that flow to an interacting
superconformal field theory (SCFT) at long distances. In section~7
we discuss the remaining open questions, most importantly,
emphasizing possible applications to EWSB, and conclude.

\newsec{The Axial Anomaly and the Energy-Momentum Tensor}

In this section we review the well-known connection between the
axial anomaly, the energy-momentum tensor, and the $R$-symmetry
current (if the latter exists). To that end, we first note that
the the anti-commutation relations, $\{Q,\bar Q\}\sim P$, imply
that the supercurrent and the energy-momentum tensor sit in the
same multiplet. The question then becomes how to write an
irreducible representation of SUSY containing the supercurrent and
the energy-momentum tensor. The simplest solution is the
Ferrara-Zumino multiplet~\FerraraPZ\foot{We adopt the following
conventions:
$\ell_{\alpha\dot\alpha}=-2\sigma^{\mu}_{\alpha\dot\alpha}\ell_{\mu}$,
$\ell_{\mu}={1\over4}\bar\sigma_{\mu}^{\dot\alpha\alpha}\ell_{\alpha\dot\alpha}$.}
\eqn\FZ{ \bar D^{\dot\alpha} \CJ_{\alpha\alphadot}=D_\alpha X~, }
where $X$ is chiral and $\CJ_{\mu}$ is real. In some cases, which
are not relevant to this paper, the FZ multiplet does not exist
\refs{\KomargodskiPC,\KomargodskiRB}.

Writing out the solution to \FZ, we find that the $\theta^2$
component of $X$ contains the trace of the energy-momentum tensor
as well as the divergence of the bottom component of $\CJ_{\mu}$.
In pure super Yang-Mills theory, $X$ is proportional to
$W_\alpha^2$. The trace of the energy-momentum tensor is
proportional to $F_{\mu\nu} F^{\mu\nu}$, while the ABJ equation
relates the divergence of the bottom component of $\CJ_{\mu}$ to
$F\tilde F$.\foot{Incidentally, this relation leads to the famous
``anomaly puzzle,'' (see \refs{\ShifmanZI\HuangTN\ArkaniHamedMJ-\YonekuraMC}, and many references therein, for a
more detailed discussion of this puzzle) since one would expect
the anomaly to be naturally one-loop exact while the beta function
has contributions from all loop orders. This apparent paradox will
not affect our discussion below in any way.}

Intuitively, it is this connection between the anomaly and the
energy-momentum tensor that allows one to say more than usual about the flow of the anomalous current. It is
relatively easy to identify the energy-momentum tensor in the IR
of a complicated flow (as long as we know what the, possibly
emergent, degrees of freedom are there). This discussion
also suggests that coupling the rigid theory to supergravity, as in \LutyQC, might
shed some light on the mapping of anomalous currents.
We will not consider supergravity in this note.

When the theory under consideration has an exact $R$-symmetry, there is a more natural representation of the supercurrent which, as we will see below, leads to a simpler description of the physics. In such a case, the conserved $R$-current transforms as the bottom component of a supercurrent multiplet that is defined by
\eqn\definingR{ \bar
D^\alphadot\CR_{\alpha\alphadot}=\chi_\alpha~. 
}
Here $\chi_\alpha$ is chiral and satisfies the usual Bianchi identitiy $D\chi=\bar D\bar\chi$, and $\CR_{\mu} $ is real. In the systems of interest to us, both the FZ multiplet~\FZ\ and the $\CR$-multiplet~\definingR\ exist. From this statement, it follows that the Bianchi identity for $\chi_\alpha$ can be solved in terms of a well-defined real superfield, $U$, and so \eqn\multiplet{
\bar D^\alphadot \CR_{\alpha\alphadot}=\bar D^2D_\alpha U~.
}

The general picture of what happens to $U$ along a flow is simple to understand. In the asymptotically free theories we will study below, $\CR$ and $U$ start out in the UV as bilinears in the various weakly-coupled superfields (with appropriate contributions of $\sim{1\over g^2}W_{\alpha}\bar W_{\dot\alpha}$ to $\CR$ and appropriate factors of $e^V$ to render $\CR$ and $U$ gauge invariant; we neglect these terms for simplicity---a more detailed recent discussion of many of the issues discussed here can be found in~\refs{\DienesTD, \MagroAJ, \KomargodskiRB,\KuzenkoAM,
\AntoniadisNJ, \DumitrescuCA, \KuzenkoNI, \ButterSC}). Indeed, solving \multiplet, one finds that for the matter superfields, $\Phi_i$, with $R$-charges, $r_i$, the expressions for $\CR_{\alpha\alphadot}$ and $U$ take the form
\eqn\Rdesc{
\CR_{\alpha\dot\alpha}=\sum_{i}\left(2D_{\alpha}\Phi_i\bar
D_{\dot\alpha}\bar\Phi^i-r_i[D_\alpha,\bar
D_\alphadot]\Phi_i\bar\Phi^i\right),
}
\eqn\Udesc{
U=\sum_{i}\left(1-{3r_i\over 2}\right)\bar\Phi^i\Phi_i~.
}
Note that the contributions to $U$ of fields with $r_i=2/3$ vanish because this is the superconformal $R$-charge for fields at the Gaussian UV fixed point.

We should elaborate on what it means to solve for $U$. Equation~\multiplet\ does not fix $U$ uniquely, but only fixes $\bar D^2D_\alpha U$. This leads to the usual supergauge ambiguity, $U\rightarrow U+\Omega+\bar\Omega$, where $\Omega$ is chiral. In writing~\Udesc\ we have discarded all such holomorphic terms. Indeed, in most cases they can be completely ignored by symmetry arguments. However, we will see cases where even if such terms are not included in the UV, adding such terms in the IR is forced on us by consistency.

It is also due to such ambiguities in solving superspace equations that we opt to use the $\CR$-multiplet rather than the
FZ-multiplet. Indeed, in the latter case one can show that ambiguities arise not only from purely holomorphic terms, but also from conserved currents (which generically exist and render the analysis harder).

Now, as we flow to the IR, we can use the $\CR$-multiplet to follow $U$. The IR is described by some SCFT, and $U$ can be described as $U\sim\Lambda^{2-d}\CO$, for some real operator, $\CO$, of dimension $d\ge2$, and some scale, $\Lambda$. We can assume the dimension of the real operator $\CO$ is $\geq2$ by unitarity and by the fact that we can remove holomorphic plus anti-holomorphic contributions.

In the case that $d>2$, $U$ formally vanishes at the IR fixed point (i.e. deep in the IR). This means that the bottom component of $\CR_{\alpha\alphadot}$ becomes the superconformal $R$-symmetry.\foot{A special case which is slightly more subtle is when the IR SCFT is approached by a marginally irrelevant operator. This can be represented by $U=\gamma J$, where $J$ is some dimension 2 operator in the IR SCFT and $\gamma$ is an anomalous dimension that {\it goes to zero} in the deep IR, as required. The general construction of this $J$ and the calculation of $\gamma$ is presented in the framework of~\GreenDA. We thank D.~Green and N.~Seiberg for helpful conversations on the matter. } In other words, the $R$-symmetry we have chosen in the UV becomes the superconformal one in the infrared. But we know this is not always the case. There could be multiple choices for the $R$-symmetry in the UV and there can also be accidental symmetries in the IR. When the $R$-current we follow doesn't flow to the IR superconformal one, then $U$ is {\it nonzero} in the IR and it flows to a certain current of dimension 2
\eqn\Uflow{
U\to{3\over2}J~. 
}
This conserved current, $J$, may be a conserved current of the full theory, or it may correspond to an accidental
symmetry of the IR fixed point.\foot{The above discussion relies on the assumption that the fixed points are conformal in addition to being scale-invariant. Whether this is always true is an open question (see~\refs{\PolchinskiDY, \DorigoniRA} for some aspects of the problem). However, in many cases of interest, like SQCD and various simple generalizations, conformality is strongly suggested by the discussion in \SeibergPQ\ and various related works. This picture has been given further reenforcement recently in \AntoniadisGN.}

Formally, $J$ has a simple description. It is just the conserved current which parameterizes the difference between the superconformal $R$-symmetry and the one in the multiplet we are following along the flow.  This can be shown by recalling that the IR superconformal theory admits the superconformal multiplet $\CR_{\mu}^{CFT}$ (i.e., the multiplet for which $\bar D^{\dot\alpha}\CR_{\alpha\dot\alpha}^{CFT}=0$ at the IR fixed point). We can write this multiplet in terms of the IR limit of \multiplet\ and the (perhaps accidentally) conserved current multiplet, $J$, as follows
\eqn\improvement{
\CR^{CFT}_{\alpha\alphadot}=\CR_{\alpha\alphadot}^{IR}-[D_\alpha,\bar
D_\alphadot] J~,\qquad U^{CFT}=U^{IR}-{3\over 2} J=0~.
}
Here $U^{IR}$ is the deep IR limit of $U$, and $J$ is the multiplet for the symmetry that mixes with the $R$-charge corresponding to $R_{\alpha\dot\alpha}$ to create the superconformal $R$-symmetry. $\CR_{\alpha\dot\alpha}^{IR}$ and $\CR_{\alpha\dot\alpha}^{CFT}$ are related via improvement transformations for the supercurrent and stress tensor.\foot{More general studies of improvements of supercurrent multiplets can be found in \refs{\KomargodskiRB,\ThomasNati}. }

We see that being able to follow the axial current relies crucially on being able to identify the superconformal $R$-symmetry. In many examples this is fixed by duality. Additionally, we have the powerful tools of~\IntriligatorJJ.

In many theories,  there are free-magnetic phases, where the IR is a Gaussian fixed point. Then the abstract discussion above takes a very simple form, since the superconformal $R$-charge is 2/3 for all the chiral fields. $U^{IR}$ is then fixed by the IR analog of \Udesc, namely \eqn\Udesci{ U^{IR}=\sum_{i}\left(1-{3r_i\over 2}\right)\bar\phi^i\phi_i~,} where the $\phi_i$ are the ``emergent'' chiral superfields at low energies and $r_i$ are their $R$-charges. The simplest example of such a theory is SQCD in the free magnetic phase, which we will now discuss in much greater detail.

\newsec{The Anomalous Current of SQCD}

In this section we will consider $SU(N_c)$ $\CN=1$ SQCD with $N_f$ in the free magnetic range, i.e. $N_c+1<N_f\le  3N_c/2$. Recall the matter content of the electric UV theory
\eqn\tableone{
\matrix{& SU(N_c) & SU(N_f)\times SU(N_f) & U(1)_R & U(1)_B \cr &
\cr  Q & {\bf N_c} & {\bf N_f\times 1}& 1-{N_c\over N_f} & 1 \cr
\tilde Q & {\bf \bar N_c} & {\bf 1\times \bar N_f}& 1-{N_c\over
N_f} & -1 }
}
A particularly interesting set of operators to try and follow is given by all the possible non-holomorphic bilinears
\eqn\Kahlerdef{
c_{i}^{j}Q^iQ^\dagger_j+\tilde c_i^j \tilde
Q^i\tilde Q^\dagger_j~,
}
where $c_i^j, \tilde c_i^j$ represent some arbitrary real numbers, and $i,j=1,...,N_f$.

The theory~\tableone\ is understood in the IR via Seiberg-duality~\SeibergPQ. The low energy degrees of freedom consist of a dual IR-free $SU(N_f-N_c)$ gauge group, $N_f$ dual quark superfields $q, \tilde q$ in the fundamental anti-fundamental representations of $SU(N_f-N_c)$, and a gauge singlet meson $N_f\times N_f$ matrix, $M$. We summarize this matter content in the following table
\eqn\tabletwo{ \matrix{&
SU(N_f-N_c) & SU(N_f)\times SU(N_f) & U(1)_R & U(1)_B \cr & \cr  q
& {\bf N_f-N_c} & {\bf \bar N_f\times 1}& {N_c\over N_f} &
{N_c\over N_f-N_c} \cr \tilde q & {\bf \bar N_f-\bar N_c} & {\bf
1\times N_f}& {N_c\over N_f} & -{N_c\over N_f-N_c}  \cr  M & {\bf
1} & {\bf N_f\times N_f}& 2-2{N_c\over N_f} & 0 }
}
Since the theory is IR-free, the natural normalization of these dual fields is to choose their kinetic terms to be canonical.

Given this picture, we would like to know how the operators in~\Kahlerdef\ are realized in the dual theory of~\tabletwo. It is difficult to answer this question exactly, since the result depends on incalculable corrections to the K\"ahler potential of the IR degrees of freedom. However, here we are only interested in knowing what the operators in \Kahlerdef\ flow to in the deep IR, where all such corrections are irrelevant.

As mentioned in the introduction, it is extremely easy to follow to the IR operators of
the form~\Kahlerdef\ that correspond to conserved currents. For example, consider $Q^iQ^\dagger_i-\tilde Q^i\tilde Q^\dagger_i$. This operator can
be identified with the bottom component of the conserved baryon
superfield \eqn\conservbaryon{ \bar D^2\left(Q^iQ^\dagger_i-\tilde
Q^i\tilde Q^\dagger_i\right)=0~. } We can immediately conclude that in the deep IR it should be matched to the baryon number current of the magnetic theory. In other words, 
\eqn\flow{ QQ^\dagger-\tilde Q\tilde Q\longrightarrow
{N_c\over N_f-N_c}\left(|q|^2-|\tilde q|^2\right)~. 
}
The numerical factor on the RHS of this equation follows from the well-known baryon charge of the magnetic quarks (see table~\tabletwo).

It is just as easy to follow some other special bilinears in the squark superfields. Indeed, all the bilinears given by linear
combinations of $QT^a Q^\dagger$ and $\tilde Q T^a\tilde Q^\dagger$ (with traceless, Hermitian, $T^a$) can be thought of as the bottom components of the non-Abelian currents associated with $SU(N_f)_L\times SU(N_f)_R$ and can thus be directly mapped to the IR (this is done via the action of these symmetries on the magnetic degrees of freedom \tabletwo).

In the space of all bilinears~\Kahlerdef\ there is, however, one linearly independent combination which is non-trivial to map to the IR. Without loss of generality, this linear combination can be chosen to be \eqn\nontrivcomb{ J_A=QQ^\dagger+\tilde Q\tilde Q^\dagger~. } This is not the bottom component of any conserved current. In fact, it is the bottom component of the anomalous axial current \eqn\anomaly{ \bar D^2 J_A\sim W_\alpha^2~. }

As we have explained in the previous sections, following anomalous currents is nontrivial. We will now see that
supersymmetry helps us bypass this problem in a simple manner.

We note that the theory~\tableone\ has a non-anomalous $R$-symmetry and so we can associate an $\CR$-multiplet to this $R$-symmetry along the flow. Using the formula~\Udesc\ and table~\tableone, we can identify $U$ in the far UV in terms of the electric quarks as \eqn\UUV{ U^{UV}=\left(-{1\over 2}+{3N_c\over 2N_f}\right)\left(QQ^\dagger+\tilde Q\tilde Q^\dagger \right),} and in the IR we can express $U$ in terms of the magnetic degrees of freedom using~\Udesci\ and the $R$-charges in table~\tabletwo\ 
\eqn\UIR{ U^{IR}=\left(1-{3N_c\over
2N_f}\right)\left(qq^\dagger+\tilde q \tilde
 q^\dagger\right)-\left(2-{3N_c\over N_f}\right)MM^\dagger.
 }
 This shows that the operator \nontrivcomb\ undergoes the following flow
 \eqn\flow{
 QQ^\dagger+\tilde Q\tilde Q^\dagger \longrightarrow {2N_f-3N_c\over
3N_c-N_f}\left(qq^\dagger+\tilde q \tilde
q^\dagger-2MM^{\dagger}\right)~. 
}
This is an exact result. In this formula~\flow\ we have chosen the mesons and magnetic quarks to be canonically normalized.\foot{Note that if one interprets~\flow\ as the action of the anomalous axial current on the IR degrees of freedom, we find that the cubic superpotenial of the magnetic theory, $W_{\rm mag}=qM\tilde q$, is invariant.}

One interesting consequence of the above discussion is that,  upon acting with $\bar D^2$ on both sides of the mapping in \flow, we find the physical relation between the electric and magnetic field strengths \eqn\WWmap{ W_{\alpha, {\rm el}}^2\longrightarrow {2N_f-3N_c\over 3N_c-N_f}W^2_{\alpha, {\rm mag}}. } This is again
an exact result.\foot{The reader may wonder how this relates to the claim $W^2_{\alpha, {\rm el}}\to-W^2_{\alpha, {\rm mag}}$ made in \IntriligatorAU\ and elsewhere. The main point is that this mapping is derived by using the holomorphic scale matching relation, which means that $W^2_{\alpha, {\rm el}}\to-W^2_{\alpha, {\rm mag}}$ is only valid modulo trivial elements of the chiral ring. By the ABJ equation, the squares of the field strengths are themselves trivial in the chiral ring of the undeformed theory and so the mapping $W^2_{\alpha, {\rm el}}\to-W^2_{\alpha, {\rm mag}}$ carries the same information as $0=0$. (The relation $W^2_{\alpha, {\rm el}}\to-W^2_{\alpha, {\rm mag}}$ has nontrivial content if the theory is deformed.) On the other hand, our result in~\WWmap\ gives the physical normalization which can be measured, for instance, by studying correlation functions at long distances. Note that the exact result also has a sign flip in the free magnetic phase, so the interpretation of one coupling growing while the other decreasing, remains.}

\subsec{Soft SUSY-Breaking}

We can immediately apply the results in~\flow\ and \WWmap\ to study the mapping of soft terms in the electric theory to soft terms in the magnetic theory. To that end, consider deforming the UV Lagrangian by adding the bottom components of the current in~\nontrivcomb\ and the electric field-strength bilinear in~\WWmap\ so that we give small squark and gaugino soft masses to the electric fields
\eqn\Ldef{
\delta\CL_{\rm el}=-m^2J_A|-{m_{\lambda}}(W^2_{\alpha, {\rm el}}+{\rm hc})|=-m^2\left(QQ^{\dagger}+\tilde Q\tilde Q^{\dagger}\right)+{m_{\lambda}}(\lambda_{{\rm el}}^2+c.c.)~,
}
where we take $m^2$ and $m_{\lambda}$ positive with $m, {m_{\lambda}}\ll\Lambda_{\rm el, mag}$, and  $\Lambda_{\rm el, mag}$ are the dynamical scales of the electric and magnetic theories respectively.

Since the soft deformations in \Ldef\ are small (compared to
$\Lambda_{\rm el, mag}$), we can treat the underlying dynamics of
the theory as supersymmetric and work in the ``probe
approximation," where the subleading $\CO(m/ \Lambda)$ and
$\CO(m_{\lambda}/\Lambda)$ corrections to the IR soft masses are
neglected.\foot{In QCD this is what one does to follow quark
masses in the chiral Lagrangian. In the chiral Lagrangian,
however, there is an incalculable overall coefficient in the
mapping. This incalculable coefficient can be expressed in terms
of the mass of the physical pion, and the probe approximation
amounts to expanding in $m_\pi\over f_\pi$. In our case, since we
solved for the mapping exactly, no incalculable coefficients
arise.} For simplicity, we also neglect possible  contributions to scalar masses squared scaling like $m_\lambda^2$. These
may be important for phenomenological applications, but we will
not discuss them here.

In this approximation, we see from \flow\ and \WWmap\ that the magnetic deformation corresponding to \Ldef\ is
\eqn\Lmagdef{
\delta\CL_{\rm mag}=-m^2\cdot{2N_f-3N_c\over
3N_c-N_f}\left(qq^\dagger+\tilde q \tilde
q^\dagger-2MM^{\dagger}\right)+m_{\lambda}\cdot{2N_f-3N_c\over
3N_c-N_f}(\lambda_{{\rm mag}}^2+c.c.)~,
}

These results agree with \refs{\ChengXG,\ArkaniHamedWC,\LutyQC}.
Our derivation shows that the ability to map soft terms follows
from the simple mapping of the electric and magnetic $R$-symmetry (the role of the $R$-symmetry was also emphasized, although from a slightly different perspective, in \LutyQC).

Note that if all the masses in the UV are positive, then, in the
IR, the magnetic squarks are tachyonic (we are in the free
magnetic phase and so $2N_f-3N_c<0$). It turns out that even the
magnetic $D$-terms and superpotential do not help to stabilize the
magnetic squarks; for example, there is an instability along the
direction $q\sim \unit$, $\tilde q=0$, $M=0$.\foot{By the equation
$q\sim \unit$, we mean that we choose the upper left
$(N_f-N_c)\times (N_f-N_c)$ block to be proportional to the unit
matrix, and the rest of the entries to be zero. The same comment
applies everywhere below.} Our approximation does not allow one to know where the theory settles.

However, it is interesting to note that we can stabilize the
dynamics by considering a simple deformation of SQCD. To see this,
consider weakly gauging baryon number with some small gauge
coupling, $g_B$. Then, it is easy to prove that there are no
instabilities which take us out of the calculable regime (as long
as $g_B$ is not too small). Indeed, one finds a vacuum with $q\sim
{m\over g_B}\unit$, $\tilde q=0$, $M=0$ and of course a similar
vacuum with $q$ interchanged with $\tilde q$. Therefore, all we
need for calculability is that $g_B$ is much larger than
$m/\Lambda$ but sufficiently smaller than all the other couplings
in the theory. This vacuum breaks the magnetic gauge symmetry and
Higgses baryon number too. The remaining non-Abelian flavor
symmetry is $SU(N_f-N_c)\times SU(N_c)\times SU(N_f)$. (Note the
color-flavor locking phenomenon. Ideas along these lines thus
present an opportunity for extending various recent studies such
as \refs{\KomargodskiMC\ShifmanKA\ShifmanXC-\EtoPJ} into the
non-supersymmetric domain.) If, on the other hand, the gauge
coupling $g_B$ is sufficiently large compared to the gauge and
Yukawa couplings of the theory, a different stable vacuum appears,
where $q\sim\tilde q\sim m$ and $M=0$. Both of these vacua will be
mentioned again briefly in the last section, motivated by some
possible phenomenological applications.

Finally, from our discussion above it is clear that we can
consider the most general set of non-holomorphic soft terms in the
UV by adding~\Kahlerdef\ and decomposing it into the soft terms
associated to the conserved currents and the anomalous current we
have discussed at length.

\newsec{The Deformed Moduli Space}

When some of the symmetries of the short-distance theory are
broken spontaneously, there are interesting subtleties in the flow
of the $U$ operator~\Udesc. In particular, the holomorphic plus
anti-holomorphic pieces of the type discussed in section 2 appear.

The deformed moduli space of $N_f=N_c$ SQCD~\SeibergBZ\ is a simple arena in which to study these ideas. Indeed, the quantum dynamics of SQCD with $N_f=N_c>2$ deforms the moduli space so that it is parameterized by baryons and mesons subject to
\eqn\deformedM{
\det M-B\tilde B=\Lambda^{2N_c}.
}
Hence, some of the UV symmetries are necessarily spontaneously broken.

We will see below that $U$ receives contributions from the
corresponding Goldstone multiplets and that requiring invariance
of $U$ under the resulting nonlinearly-realized symmetries both
necessitates the inclusion of holomorphic plus anti-holomorphic
corrections to $U$ that are quadratic in the Goldstone multiplets
and, simultaneously, fixes their mixing with $U$ exactly. We will
also see a vacuum in which this ambiguity is not fixed by
symmetries.

Even though the global symmetries are spontaneously broken, it is
still straightforward to follow conserved currents to the
IR.\foot{There could, however, be some complications. In addition
to the one already mentioned, analogous to the complication in following the baryon current in
QCD, there are also exotic cases when the ordinary linear
multiplets are not globally well defined, see~\DumitrescuCA\ and
references therein.}

The anomalous current, is, of course, harder to follow. To proceed, we consider the following highly symmetric vacuum satisfying~\deformedM\ \eqn\vactwo{ M=0~,\qquad B=\tilde B=\Lambda^{N_c}~. } This vacuum breaks the symmetry according to $SU(N_f)_L\times SU(N_f)_R\times U(1)_B\times U(1)_R\hookrightarrow SU(N_f)_L\times SU(N_f)_R\times U(1)_R.$ The massless fluctuations in this vacuum are the meson matrix $\delta M$ and the Goldstone superfield, $\delta b$, associated with $U(1)_B$ breaking.

We are interested in finding the low energy limit of the axial current, $J_A=QQ^{\dagger}+\tilde Q\tilde Q^{\dagger}$. Noting that all the chiral fields have vanishing $R$-charge, we use \Udesci, and immediately find that, up to holomorphic plus anti-holomorphic pieces, $U=\delta M\delta M^{\dagger}+\delta b\delta b^{\dagger}$. Note, however, that this operator is not invariant under the non-linear imaginary shift symmetry of $\delta b$. Therefore, we must replace $\delta b\delta b^{\dagger}\to {1\over2}(\delta b+\delta b^{\dagger})^2$ and we conclude that
\eqn\UIR{
QQ^{\dagger}+\tilde Q\tilde
Q^{\dagger}\longrightarrow\Tr\left(\delta M\delta
M^{\dagger}\right)+{1\over2}(\delta b+\delta b^{\dagger})^2~. 
}
We see that the addition of a purely holomorphic and anti-holomorphic piece quadratic in the Goldstone multiplet is forced on us. The answer~\UIR\ is exact in the deep IR, in particular, there are no further holomorphic ambiguities.\foot{The $Z_2$ interchange symmetry acting on the UV degrees of freedom as $Q\leftrightarrow\tilde Q$, with an appropriate action on the vector superfield, rules out the appearance of the linear term $\delta b+\delta b^{\dagger}$. The remaining linearly realized symmetries also force holomorphic contributions in $\delta M$ to vanish.}

Unlike the discussion of the previous section, just adding a soft deformation in the UV, $\delta
\CL=-m^2\left(QQ^{\dagger}+\tilde Q\tilde Q^{\dagger}\right)$, is enough to end up with a stable vacuum in the IR. Equation~\UIR\ shows that all the meson fluctuations are massive, the real part of the baryon is massive as well, and the imaginary part is the ordinary Goldstone boson for $U(1)_B$ breaking in this non-SUSY vacuum.

One can also consider adding a soft gaugino mass in the UV. Even
though there are no gauge fields in the IR, this affects the
infrared in a nontrivial way at leading order in the gaugino mass.
We can hit~\UIR\ from both sides with $\bar D^2$. On the left hand
side we get the usual fields strength squared operator from the
ABJ equation, while on the right hand side most terms vanish by
the free equations of motion (neglecting irrelevant corrections
from the K\"ahler potential). However, when we hit $(\delta
b^\dagger)^2$, we get $\sim \bar\psi_b\bar\psi_b$. In other words,
up to an order one coefficient, ${1\over
8\pi^2}W_\alpha^2\longrightarrow \bar D_\alphadot b^\dagger \bar
D^\alphadot b^\dagger$. As a result, a gaugino mass in the UV
manifests itself in the IR via a mass term for the fermionic
partner of the Goldstone boson.\foot{Note that this fermionic mass
term is enhanced by a loop factor compared to the gaugino mass.
This could have some interesting phenomenological applications,
because it is usually hard to generate large fermionic masses
compared to scalars in the same multiplet. One possible connection
to phenomenology could thus be through the problems revolving
around the $\mu$-term.}

The deformed moduli space has another vacuum with an enhanced symmetry
\eqn\vacone{
M=\Lambda\unit~,\qquad B=\tilde B=0~.
}
The symmetry breaking here is $SU(N_f)_L\times SU(N_f)_R\times U(1)_B\times U(1)_R\hookrightarrow SU(N_f)_V\times U(1)_B\times U(1)_R$.
In this vacuum, the massless fluctuations are the traceless mesons $\delta M$ in the $Adj_{(0,0)}$ representation, and the baryons, $\delta B$ and $\delta\tilde B$, in the $0_{(\pm N_c,0)}$ representation.

Repeating the mapping of the axial current, we again find that some holomorphic terms in the mesons are necessarily induced with known coefficients. However, we now have an ambiguous chiral singlet operator of the form $\delta B\delta\tilde B$, whose mixing with $U$ we cannot fix. We therefore add it with an unknown coefficient, $c$
\eqn\flowdeformedi{
QQ^\dagger+\tilde Q\tilde Q^\dagger\longrightarrow \half Tr\left(\delta M+\delta M^\dagger\right)^2 + \delta B\delta B^\dagger+\delta \tilde B\delta \tilde B^\dagger+ c\left(\delta B\delta\tilde B+c.c.\right)~.
}
This ambiguity prevents us from making exact statements about the nature of this vacuum when we softly deform the theory in the UV.

The case of $N_f=N_c=2$ might be interesting for model building,
so we comment on it too. In the most symmetric vacuum, one finds
the symmetry breaking pattern $SO(6)\hookrightarrow SO(5)$. The
fluctuations are in two five-dimensional representation of
$SO(5)$. This symmetry precludes any linear terms in the
fluctuations from appearing in the map and so the symmetric point
remains an extremum upon softly deforming the theory in the UV by
the bottom component of the axial current. With identical tools to
those we have used above, one also finds that all the partners of
the Golstone bosons are stabilized and hence the most symmetric
point is a local minimum. There is no ambiguity in quadratic
holomorphic terms, and all the masses are calculable, as in all
the examples we have studied besides~\vacone.

\newsec{Kutasov Duality}
In the above sections we considered theories with only one non-conserved current in the UV---the current corresponding to the anomalous symmetry. In this section, we will analyze theories with more non-conserved currents in the UV. A simple example is given by adjoint SQCD with a superpotential for the adjoint, $X$. In such a case there are two independent non-conserved currents in the UV---the one corresponding to $U$, which sits in the same multiplet as the (unique) non-anomalous $R$-symmetry, and a non-anomalous current which is explicitly broken by the
superpotential.

The theories we will discuss in this section were studied in \refs{\KutasovVE,\KutasovNP,\KutasovSS} and have the following particle content and symmetries
\eqn\tablethree{\matrix{& SU(N_c)
& SU(N_f)\times SU(N_f) & U(1)_R & U(1)_B \cr & \cr Q & {\bf N_c}
& {\bf N_f\times 1}& 1-{2\over k+1}{N_c\over N_f} & 1 \cr \tilde Q
& {\bf \bar N_c} & {\bf 1\times \bar N_f}& 1-{2\over k+1}{N_c\over
N_f} & -1 \cr X & {\bf N_c^2-1} & {\bf 1\times 1}& {2\over k+1} &
0 } } 
The superpotential for the adjoint has the form $W=s_0\Tr(X^{k+1})$ and breaks the symmetry associated with the
non-anomalous current 
\eqn\specialcomb{
J_X={N_c\over
N_f}\left(QQ^\dagger+\tilde Q\tilde Q^\dagger\right)-XX^\dagger, 
}
where $\bar D^2J_X\sim s_0\Tr(X^{k+1})$. The other non-conserved
current is just the anomalous (and, for $k>2$, broken by the superpotential) current, $U$, associated with the $\CR$-multiplet
\eqn\anomcomb{ U^{UV}=\left(-\half+{3\over k+1} {N_c\over
N_f}\right)\left(QQ^\dagger+\tilde Q\tilde
Q^\dagger\right)+\left(1-{3\over
 k+1}\right)XX^\dagger.
}

In what follows, we will focus mostly on the free magnetic phase (${N_c\over k}<N_f<{2N_c\over 2k-1}$) where the dual description is a weakly coupled theory with the following massless fields
\eqn\tablefour{
\matrix{& SU(kN_f-N_c) & SU(N_f)\times
SU(N_f) & U(1)_R & U(1)_B \cr & \cr  q & {\bf {kN_f-N_c}} & {\bf
\bar N_f\times 1}& 1-{2\over k+1}{k N_f-N_c\over N_f} & {N_c\over
kN_f-N_c} \cr \tilde q & {\bf \overline{kN_f- N_c}} & {\bf 1\times
N_f}& 1-{2\over k+1}{k N_f-N_c\over N_f} & -{N_c\over k N_f-N_c}
\cr Y & {\bf (kN_f-N_c)^2-1} & {\bf 1\times 1}& {2\over k+1} & 0
\cr M_j & {\bf 1} & {\bf N_f\times\bar N_f} & 2-{4\over
k+1}{N_c\over N_f}+{2\over k+1}(j-1) & 0}
}
and the following superpotential
\eqn\Wmag{
W_{\rm mag}=-{s_0\over k+1}{\rm Tr}\
Y^{k+1}+{s_0\over\mu^2}\sum_{j=1}^kM_j\tilde q Y^{k-j}q~.
}

Let us now consider the mapping of the currents of the theory to the IR. The mapping of the conserved currents proceeds trivially as before. The mapping of the $U$ operator follows from our general discussion above with the non-trivial result that
\eqn\flowkutasov{\eqalign{ U^{IR}&= \left(-\half+{3\over
k+1}{kN_f-N_c\over N_f}\right)   \left(qq^\dagger+\tilde q\tilde
 q^\dagger\right)+\left(1-{3\over k+1}\right)YY^\dagger\cr&+\sum_j \left(-2+{6\over k+1}{N_c\over N_f}-{3(j-1)\over k+1}\right)M_jM_j^\dagger~.
 }}

While we are able to use our methods to map all the conserved currents and the non-conserved operator~\anomcomb, there is one current whose mapping we cannot fix---namely that of $J_X$. Being able to follow such an operator would amount, via  $\bar D^2J_X\sim s_0\Tr(X^{k+1})$, to following $s_0\Tr(X^{k+1})$, but since the latter vanishes in the chiral ring this is not straightforward (any formula obtained from chiral ring relations cannot be trusted since it contains the same information as $0=0$).\foot{By matching chiral primaries, we can, however, show that the charge of $Y$ under $J_X$ is the same as that of $X$.}

\newsec{Conformal Theories}
In section 2 we described the flow of $U$ when the IR is given
by some general SCFT, but so far we have focused mostly on theories
with a free IR description. In this section we will briefly consider theories with an interacting IR fixed point.

Let us start from SQCD in the conformal window $3N_c/2\le N_f<3N_c$. Now, the non-anomalous $R$-symmetry of the ultraviolet~\tableone\ becomes the superconformal one in the IR, and so in the deep IR $QQ^\dagger+\tilde Q\tilde Q^\dagger$ flows to zero. But we would like to say a little more. At the onset of the conformal window $N_f=3N_c/2$, the free fixed point in the IR is approached logarithmically, due to a marginally irrelevant operator. This means that $QQ^\dagger+\tilde Q\tilde Q^\dagger$ flows to zero in the deep IR logarithmically too. (Indeed, the right hand side of~\flow\ vanishes upon substituting $N_f=3N_c/2$.) However, in the bulk of the conformal window, the fixed points are approached by strictly irrelevant operators,\foot{One can argue that this is true as follows. The conventional wisdom about the conformal window of SQCD is that in the bulk of it there are no accidental symmetries in the infrared. However, marginally irrelevant operators must violate some of the symmetries of the SCFT~\refs{\KolZT,\GreenDA}. But since the RG flow preserves all the symmetries, we conclude that the conformal point is not approached via marginally irrelevant operators.} and so $QQ^\dagger+\tilde Q\tilde Q^\dagger$ flows to some operator of dimension $>2$ in the IR SCFT, divided by an appropriate power of the strong scale, $\Lambda$.

Let us see what this implies for soft deformations of the theory. Suppose we softly deform the theory in the ultraviolet by $\delta \CL=-m^2\left(QQ^\dagger+\tilde Q\tilde Q^\dagger\right)$. Then, in the bulk of the conformal window, all the effects of this deformation in the infrared (say at energy scales of order $m$) are suppressed by powers of $\Lambda$, which can be thought of as an ultraviolet cutoff at low energies. Therefore, unlike the examples we have studied in the free magnetic phase, here the effects of a deformation at the scale $m$ in the UV may become important only at much lower energy scales. For instance, this scale would be $~m^2/\Lambda$ if the first term appearing in $U$ is a real operator in the SCFT of dimension $3$ divided by $\Lambda$. This scenario can be thought of as a very close relative of the phenomenon that non-BPS operators obtain positive anomalous dimensions in SCFTs, which then lead to suppressed effects of non-SUSY deformations.\foot{Such a setup may even lead to accidental SUSY and may be of phenomenological interest. See the nice recent discussion in~\SundrumGV\ and references therein.}

A more interesting example to consider is adjoint SQCD without a superpotential. Some of the fields in the IR decouple, and allow us to write an explicit expression for their contribution to $U^{IR}$. The matter content and representations of this theory are
\eqn\tablefive{ \matrix{& SU(N_c) & SU(N_f)\times SU(N_f) & U(1)_R & U(1)' & U(1)_B \cr & \cr Q & {\bf N_c} & {\bf N_f\times 1}& 1-{2N_c\over3N_f} & 1 & 1 \cr \tilde Q & {\bf \bar N_c} & {\bf
1\times \bar N_f}& 1-{2N_c\over3N_f} & 1 & -1 \cr X & {\bf
N_c^2-1} & {\bf 1\times 1}& 2/3 & -1 & 0 } 
}
Associated with the $R$-symmetry in~\tablefive\ one finds the axial anomaly operator in the UV, $U^{UV}=\left(-{1\over2}+{N_c\over N_f}\right)\left(QQ^\dagger +\tilde Q\tilde Q^\dagger\right)$. We would like to find the IR end point of the flow for this operator.

The procedure summarized in~\Uflow\ and~\improvement\ instructs us
to identify the superconformal $R$-symmetry in the IR, and, once
this is done, the end point of the flow of
$\left(-{1\over2}+{N_c\over N_f}\right)\left(QQ^\dagger +\tilde
Q\tilde Q^\dagger\right)$ is determined: up to an overall factor
of $3/2$, it is simply the global symmetry current operator that
makes up for the difference between the $R$-symmetry
in~\tablefive\ and the superconformal one (which can be determined
from a-maximization in this case).

As has been discussed in great detail
in~\refs{\KutasovIY,\IntriligatorMI}, for small enough values of
$N_f/N_c$ there are also free fields in the low energy SCFT. For
instance (at large $N_c$), if $N_f/N_c<(3+\sqrt7)^{-1}$, then
$M_0=Q\tilde Q$ becomes free. Upon lowering $N_f/N_c$ further,
more and more mesons of the form $M_i=QX^i\tilde Q$ become free.
Their superconformal $R$-charge is therefore corrected to be $2/3$
and we can immediately use~\Uflow\ to determine how they appear in
the low energy expression of the operator
$\left(-{1\over2}+{N_c\over N_f}\right)\left(QQ^\dagger +\tilde
Q\tilde Q^\dagger\right)$ \eqn\lowenergy{\eqalign{
\left(-{1\over2}+{N_c\over N_f}\right)\left(QQ^\dagger +\tilde
Q\tilde Q^\dagger\right)\longrightarrow& \sum_{j=0}^{P(N_f/N_c)}
\left(1-{3R(M_j)\over
2}\right)M_jM_j^\dagger+\cdots\cr&=-\sum_{i=0}^{P(N_f/N_c)}
\left(j+2-2{N_c\over N_f}\right)M_jM_j^\dagger+\cdots~. }} Here
the $\cdots$ stand for an operator (which is also a global current
according to~\Uflow) in the interacting SCFT module. We have only
displayed the contributions from the free fields, because these
are the ones that can be represented explicitly in terms of some
well-defined degrees of freedom. Also, $P(N_f/N_c)$ is defined to
be the number of free fields for the given value of $N_f/N_c$.
This function can be deduced from a-maximization.

Note that~\lowenergy\ implies that if we softly deform the UV
theory by adding a mass squared for the electric scalars of the
form, $\delta\CL=-m_{UV}^2(QQ^\dagger+\tilde Q\tilde Q^\dagger)$,
the free fields in the infrared acquire a leading-order mass
squared of the form $m^2_j= \left({N_f\over
N_c-N_f/2}\right)\left(2N_c/N_f-2-j\right)m^2_{UV}$. For instance,
when $M_0$ becomes free, $N_f/N_c\leq(3+\sqrt7)^{-1}$, such a soft
deformation in the UV would stabilize it at the origin.

\newsec{Discussion and Open Questions}

In this note, we have described a simple way to follow anomalous
currents along the RG flow. In the context of Seiberg duality,
this extends the map of operators to anomalous (non-chiral)
current multiplets. We have also seen that there are some simple
results for theories whose low energy description is given in
terms of an interacting SCFT. Beyond the general interest in
understanding the maps of different operators under complicated RG
flows, our study could be of phenomenological relevance in
supersymmetric models of compositeness, and most obviously in
models of composite electroweak symmetry breaking.

\eqn\tablecomp{ \matrix{ &               & SU(N_c)   & SU(2)_L &
U(1)_Y \cr \cr & H            & {\bf N_c} & {\bf 2}& {1\over2} \cr
& \tilde{H}  & {\bf {\bar N}_c} & {\bf 2}& -{1\over2} \cr
&\Phi_{i=1\ldots N_c-1 } & {\bf N_c} & {\bf 1} & {1\over2} \cr
&\tilde{\Phi}_{i=1\ldots N_c-1 } & {\bf {\bar N}_c} & {\bf 1} &
-{1\over2} \cr\cr } }

It is worth presenting a simple example that illustrates how these
results might be applied. Consider the model for a composite Higgs
sector shown in Table~\tablecomp, which consists of an SQCD theory
with $N_{f}=N_{c}+1$ and $N_{c}\geq3$, and $U(1)_B$ being
identified with hypercharge.

The confined phase of this theory has $N_{f}=N_{c}+1$
baryon/antibaryon pairs,
$B,\tilde{B}\equiv(h,\tilde{h},\phi_{i=1\ldots
N_{c}-1},\tilde\phi_{i=1\ldots N_{c}-1})$ and also mesons $M$ of
zero hypercharge, which can be identified as follows
\eqn\UVspectrum{ M\equiv\left\{ \matrix{ h_{i=1\ldots N_{c}-1} = &
(H\tilde{\Phi}_{i})\cr \tilde{h}_{i=1\ldots N_{c}-1}  = &
(\tilde{H}\Phi_{i})\cr {\cal{T}}+\eta =& (H\tilde{H})\cr \eta_{ij}
=& (\Phi\tilde{\Phi}) } \right. } One also adds the usual
tree-level superpotential \eqn\Wconf{
W^{(conf)}=\,\tilde{B}MB-\Lambda^{3-N_{f}}\det M~, } where we
suppress flavor indices.

The nice feature of this model is that the hypercharge is
identified with $U(1)_B$ which is therefore gauged. As we
mentioned in Section~3, when $g_{Y}$ is of order unity and
sufficiently large with respect to the Yukawa coupling, one
obtains a minimum in which the baryons and antibaryons get VEVs of
order $m$, but the mesons' VEVs are zero. This can naturally break
$SU(2)_L\times U(1)_Y\hookrightarrow U(1)_{QED}$. Note that the
$SU(2)_L$ triplet does not obtain a VEV. This is
phenomenologically desirable.

An obvious technical issue that needs to be taken care
of is vacuum alignment, namely, forcing the theory to break
the global symmetries in the required fashion, and avoiding other points on the
Goldstone manifold. For this it is promising to consider explicit
breaking of flavor in the SUSY breaking operators. Then only the
baryon/antibaryon pair with the most negative mass-squared in the
IR gets a VEV, while the remaining modes are all massive. In
particular all the erstwhile Goldstone modes associated with the
broken global flavor symmetries are stabilized (except for the
$R$-axion, which can be lifted by other means).

This kind of set-up seems very advantageous. The scale of
electroweak symmetry breaking is naturally of order the SUSY
breaking parameters. The stable EWSB minimum appears
automatically, perhaps without some of the complications
conventional MSSM electroweak symmetry breaking entails.

Note that there are many alternative possibilities: for example by
taking the minima with $B\neq 0$ and $\tilde{B}=M=0$ that appear
when $g_B$ is smaller than the Yukawa coupling (up to some
numerical constant), and embedding both $h$ and $\tilde{h}$ in the
baryons $B$ (and of course dropping the $Y\equiv B$
identification), one can easily construct models without triplets. Development of these and similar ideas will be the subject of
future work~\refs{\Future}.

Our results may also have applications in the context of gauge
mediation and more general supersymmetric technicolor model
building as well. (See~\DineZA, and for some more modern work on
the subject see~\refs{\DobrescuGZ\LutyFJ-\FukushimaPM} and
references therein.) These mappings of operators could also be
relevant in attempts to interpret the MSSM as the magnetic, low
energy, theory of some completely different degrees of freedom.
(This idea is due to~\SeibergPQ\ and a relation to coupling
constant unification was pointed out recently in~\AbelTX.)
Perhaps, some of the applications for particle physics would
require one to understand the map beyond the ``probe
approximation.'' Moving beyond this approximation would also be an
interesting theoretical question to investigate.

\bigskip
\bigskip
 \centerline {\bf Acknowledgments}
\noindent We would like to thank O.~Aharony, I.~Antoniadis,
D.~Green, and N.~Seiberg for helpful conversations. S.A.A.
acknowledges support from the Leverhulme Trust. M.B. would like to
thank the high energy theory groups of the IFT at the Universidad
Aut\'onoma de Madrid, the \'Ecole Normale Sup\'erieure, and
NORDITA for their wonderful hospitality during various stages of
this project. The work of M.B. is funded by the ERC Advanced Grant
226371 and the contract PITN-GA-2009-237920. Z.K. gratefully
acknowledges support from DOE grant DE-FG02-90ER40542. Z.K. would
also like to thank the theory group at CERN for its warm
hospitality during the final stages of this project. Any opinions,
findings, and conclusions or recommendations expressed in this
material are those of the authors and do not necessarily reflect
the views of the funding agencies.

\listrefs
\end